\documentclass[aps,prb,10pt,twocolumn,showpacs,showkeys,superscriptaddress]{revtex4-1}
\usepackage{amssymb,graphicx,color,amsmath,xfrac,bm,stmaryrd,trimclip}
\usepackage[mathletters]{ucs}
\usepackage[utf8x]{inputenc}
\usepackage{amsmath}
\usepackage[normalem]{ulem}
\definecolor{lightblue}{rgb}{0.17,0.39,1}

\usepackage{multirow}
\usepackage{blindtext}

\newcommand{\DTO}{$\text{Dy}_2\text{Ti}_2\text{O}_7$}

\begin{document}

\title{Magnetic monopole relaxation effects in spin ice \DTO} 

\author{R. Edberg}
\affiliation{Department of Physics, KTH Royal Institute of Technology, SE-106 91 Stockholm, Sweden}

\author{A. Khansili}
\affiliation{Department of Physics, Stockholm University, SE-106 91 Stockholm, Sweden}

\author{I. M. Fjellv{\aa}g}
\affiliation{Centre for Materials Science and Nanotechnology, Department of Chemistry, University of Oslo, Norway}

\author{L. {{\O}}rduk Sandberg}
\affiliation{Niels Bohr Institute, University of Copenhagen, Denmark}

\author{P. P. Deen}%
\affiliation{Nanoscience Center, Niels Bohr Institute, University of Copenhagen, Universitetsparken 5, DK-2100 Copenhagen {\O}, Denmark}
\affiliation{European Spallation Source ERIC, 22363 Lund, Sweden}

\author{K. Lefmann}%
\affiliation{Niels Bohr Institute, University of Copenhagen, Denmark}

\author{P. Henelius}%
\affiliation{Department of Physics, KTH Royal Institute of Technology, SE-106 91 Stockholm, Sweden}
\affiliation{Faculty of Science and Engineering,  \r{A}bo Akademi University, \r{A}bo, Finland}

\author{A. Rydh}
\affiliation{Department of Physics, Stockholm University, SE-106 91 Stockholm, Sweden}

\date{\today}

\begin{abstract}

Spin ice compounds enable the exploration of the dynamics of magnetic monopoles in condensed matter systems. In this study,
we use ac calorimetry to probe the dynamical response of the heat capacity of the classical spin-ice compounds \DTO\ at low temperatures ($0.5-5\textup{ K}$). Using frequencies of 0.01-500~Hz, we find a strong frequency dependence in the measured heat capacity and are able to study thermal relaxation effects on the corresponding timescales. The relaxation time $\tau$ is determined from the frequency dependence of the heat capacity as the characteristic frequency below which the heat capacity saturates. The extracted $\tau$ shows a divergent behavior below 1\,K reaching $\sim$6\,s at 0.65\,K, similar to the relaxation time seen in previous studies \cite{Snyder2004, Klemke2011, Yaraskavitch2012}. Corresponding specific heat shows a maximum around this temperature. Performing dynamic Monte Carlo simulations, we verify that the specific heat frequency response has its origin in the slow magnetic monopole dynamics indigenous to spin ice. We find a timescale of $20\textup{ ms}$ per Monte Carlo step at $4\textup{ K}$ in contrast to $2.5\textup{ ms}$  mentioned in previous studies by other techniques \cite{JaubertMCtime}.

\end{abstract}

\maketitle

\section{Introduction}
In highly frustrated magnetic compounds the dynamic and static properties are frequently intertwined. The slow dynamics of excitations such as magnetic monopoles in the spin ice family of materials~\cite{jaubert09,JaubertMCtime,Revell13}, the long timescales observed in spin glasses \cite{Binder86,Mydosh15} or the motion of clusters in dilute magnets \cite{Biltmo2012}  are examples where the experimental time window may influence the outcome of measurements of thermodynamic properties traditionally considered static, such as the specific heat. Deducing the material properties becomes challenging both experimentally and theoretically and this field has given rise to many longstanding questions such as the nature of the spin glass ground state \cite{Binder86,Mydosh15}. 

Experimentally the dynamic properties are probed using for example out-of-phase magnetic susceptibility \cite{Snyder2004, Yaraskavitch2012, Revell13}, inelastic neutron scattering \cite{MorrisCv}, or muon spin resonance \cite{Lago2007}. In this study, we probe the low-temperature dynamic properties of the classical spin ice compound  \DTO\ (DTO) using ac calorimetry. This method offers several advantages; compared to inelastic neutron scattering the need for experimental facilities is modest and slower timescales can be probed as neutron scattering experiments typically probe timescales in the GHz to THz range. According to Griffiths's theorem \cite{Griff68} the specific heat is independent of the sample shape in the absence of an applied magnetic field, and a potentially challenging demagnetization analysis \cite{Tweng17} is avoided, compared to magnetic susceptibility. Furthermore, it is possible to perform the measurements on small samples of sub-mm size, which may be an advantage if it is difficult to synthesize large samples, or if the dynamical properties are strongly size-dependent.

While ac susceptibility measurements in the $10^{-4}-10^{4}$ Hz frequency range have become a standard complement to static susceptibility investigations, the vast majority of specific heat measurements are performed as static relaxation measurements.  The reason for this is partially due to instrumentation. The idea of ac calorimetry dates back many years \cite{Sullivan68, GMELIN19971}, but only recently has the method matured to yield absolute accuracy with the development of membrane-based calorimeters and resistive thin-film thermometry \cite{TAGLIATI201166, TAGLIATI2012RSI}. This new level of accuracy now allows for the determination of the variable frequency and dynamic heat capacity. The timescales which can be accessed using ac calorimetry, combined with the importance of specific heat as a probe of central importance in magnetic systems make this an ideal tool to investigate highly frustrated magnetic systems with slow dynamics. 

Here, we focus on the classic spin-ice system DTO, since the low-temperature specific heat in these materials has been the subject of intense recent research \cite{Castelnovo2012, Bram20}. Besides the broad spin-freezing peak in specific heat at $\sim$1\,K, a low-temperature ordering transition is theoretically predicted \cite{Melko01} to occur at $\sim$0.2\,K. Experimentally, no signs of this transition had been detected in Refs.~\cite{Ramirez, KlemkeDTOCV, Higashinaka, MelkoGingras}. However, a relaxation specific heat study found clear signs of an impending upturn appearing at 0.5\,K \cite{Pomaranski2013}. The investigation considered very long relaxation processes of more than 100 hours per measurement point at the lowest temperature (0.34 K). Subsequent neutron investigations with in situ equilibration times of several weeks did not reproduce the findings \cite{Giblin2018}, and theoretical studies \cite{henelius16} could not match the experimental upturn in the specific heat.

Ac calorimetry offers the possibility to study small samples that can be kept in thermal equilibrium at high frequencies. In analogy with magnetic susceptibility, the response from frequency oscillation can be described with a complex heat capacity. The real and imaginary parts of the complex heat capacity have many of the same interpretations as other response functions and are related by the Kramers-Kronig relations \cite{Baur1998,complexC_mario2019}. In this work we measure the real part of the complex heat capacity, which is commonly termed the dynamic heat capacity. With smaller than previously measured samples and with use of this complimentary method we hope not only to extend the knowledge of low-temperature properties of spin ice but also to shed further light on complex heat capacity measurements in the context of magnetically frustrated compounds.     

Our study finds a strong frequency dependence in the low-temperature (0.5-5\,K) dynamic heat capacity of DTO in the frequency range of 0.01-500 Hz. We do not observe any signature of the upturn below 0.5\,K as reported in the previous study \cite{Pomaranski2013}. Dynamic Monte Carlo studies of the general dipolar spin ice model \cite{Yavo08} match the experimental results well. This suggests that, at low temperatures, the main dynamic effects arising from the motion of the magnetic monopole excitations effectively capture the temperature and frequency dependence of the measured heat capacity.

\section{Experimental Method}
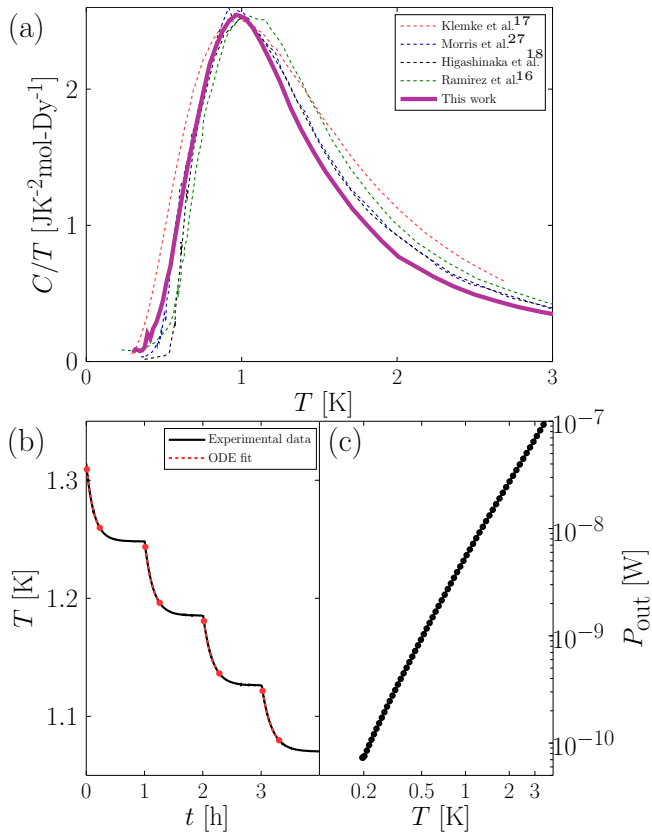
\begin{figure}[t!]
	\centering
	\vspace{80mm}
	\begin{picture}(100,100)
	\put(-71,0){\includegraphics[width=1\linewidth]{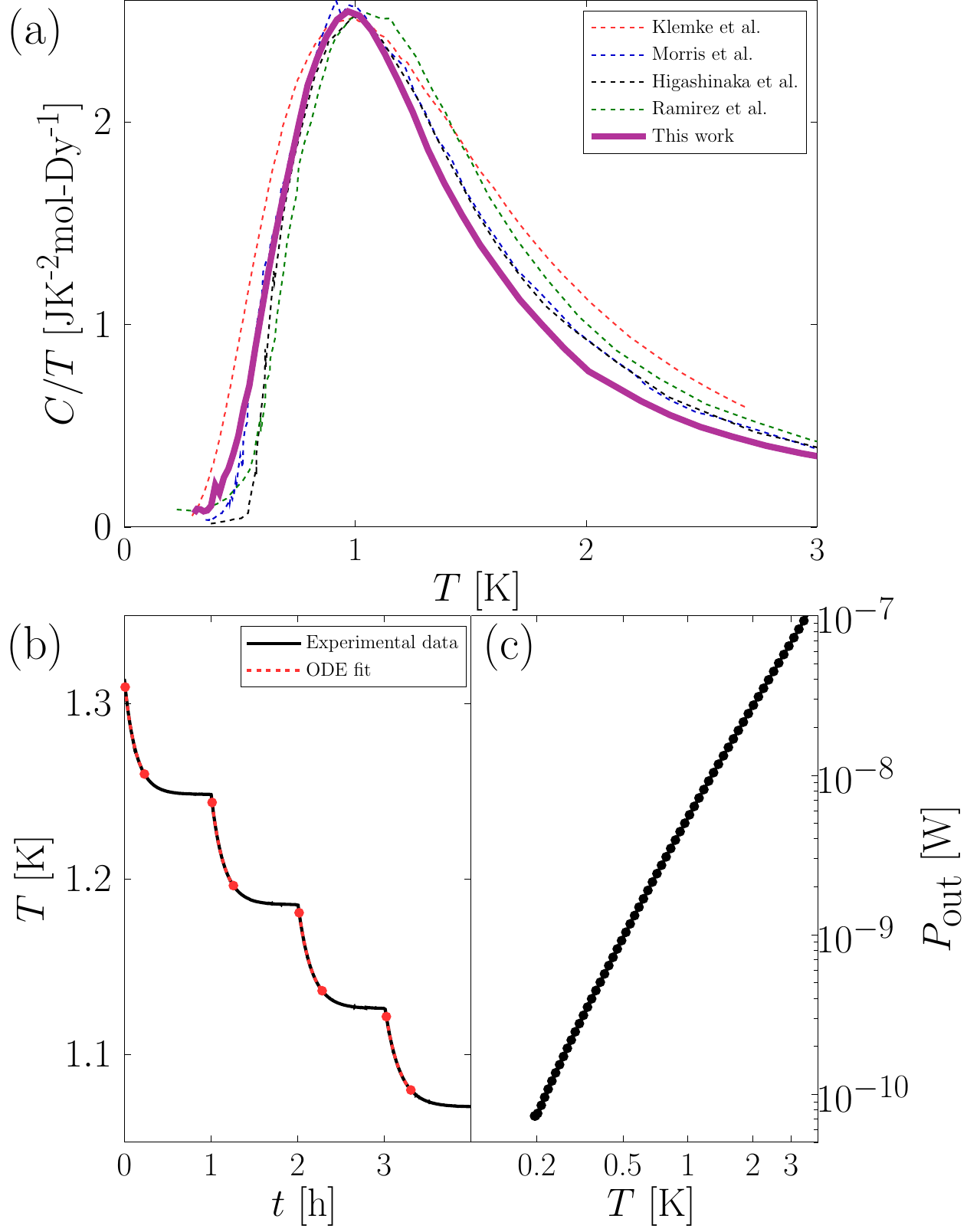}}
	\put(122,305){\tiny \cite{KlemkeDTOCV}}
	\put(120,298){\tiny \cite{MorrisCv}}
	\put(126.9,292.2){\tiny \cite{Higashinaka}}
	\put(123,284){\tiny \cite{Ramirez}}
	\end{picture}
	\caption{Heat capacity and thermal relaxation properties of a DTO powder sample cold pressed with silver powder. (a) Our heat capacity values compared with four data sets from the literature. (b) Thermal relaxation to which the solution of Eq.~(\ref{eq:heatCapacityODE}), (ODE fit), is fitted to determine $C(T)$ shown in red. (c) Experimentally measured $P_\textup{out}$ as function of temperature, used in Eq.~(\ref{eq:heatCapacityODE})}
	\label{fig1:dtomassive}
\end{figure}
The powder of DTO was synthesized and the synthesis route was a modified Pechini method (see Ref.~\cite{Sunde2016}). XRD for the powder sample shows a phase pure sample of \DTO\, with no impurity phases. To increase the thermal conductivity, the powder samples are either mixed in a 1:1 volumetric ratio with Apiezon N grease, or cold pressed in a 1:1 volumetric ratio with $99.99\%$ pure, $3~\mu\textup{m}$ Ag powder. The cold pressing method is generally used in the spin-ice literature because it gives large sturdy samples and Ag has a negligible contribution to the heat capacity \cite{Ramirez}. However, for small samples, the surface tension of Apiezon N grease is enough to give the desired rigidity, and its contribution to the heat capacity is also negligible at low temperatures. We therefore believe that both methods work equally well for the intended purpose. The samples are mounted on a Si\textsubscript{3}N\textsubscript{4} membrane calorimeter, which is connected to a thermal bath with a temperature of approximately $20\textup{ mK}$. The exact measurement procedure is described elsewhere \cite{TAGLIATI201166}. In our study, we measure a variety of powder samples of different sizes and shapes. 

\section{Results}

First, we use thermal relaxation calorimetry (see Appendix~\ref{sec:ThermalRelax}) to measure the specific heat of the powdered samples. This allows us to compare the specific heat with the literature and establishes the purity and correct measurement setup conditions. The results are shown in Fig.~\ref{fig1:dtomassive}, where panel (a) shows the specific heat obtained using the thermal relaxation method. Figure~\ref{fig1:dtomassive}(b) shows the thermal relaxation and the fit to Eq.~(\ref{eq:heatCapacityODE}) in red. Figure~\ref{fig1:dtomassive}(c) shows the measured $P_{\mathrm{out}}$ as a function of temperature used in Eq.~(\ref{eq:heatCapacityODE}) to obtain the heat capacity.
The measured heat capacity matches well with previous observations \cite{Ramirez,KlemkeDTOCV}. The overall scale has been adjusted so that the intensity of the $1\textup{ K}$ peak matches with the literature because the mass of the sample could not be determined with sufficient accuracy. The measured heat capacity is slightly lower at higher temperatures compared with the literature, and we believe that this is due to phonon contributions which can change with the mixing ratio and the sample holder used. We conclude that the setup has adequate resolution and is working correctly.

To probe the dynamics of the spin-ice system, we turn to ac calorimetry (see Appendix~\ref{sec:ModulationCal}). In Figure~\ref{fig2:DTO_S1}(a) we show heat capacity measurement obtained using modulation calorimetry for a DTO powder sample with an estimated mass of $30\,\textup{ng}$. Several measurements are performed to probe different frequencies ranging from 0.1~Hz to 200~Hz ($f$-range 1-5).

\begin{figure}[t!]
	\centering
	\includegraphics[width=0.9\linewidth]{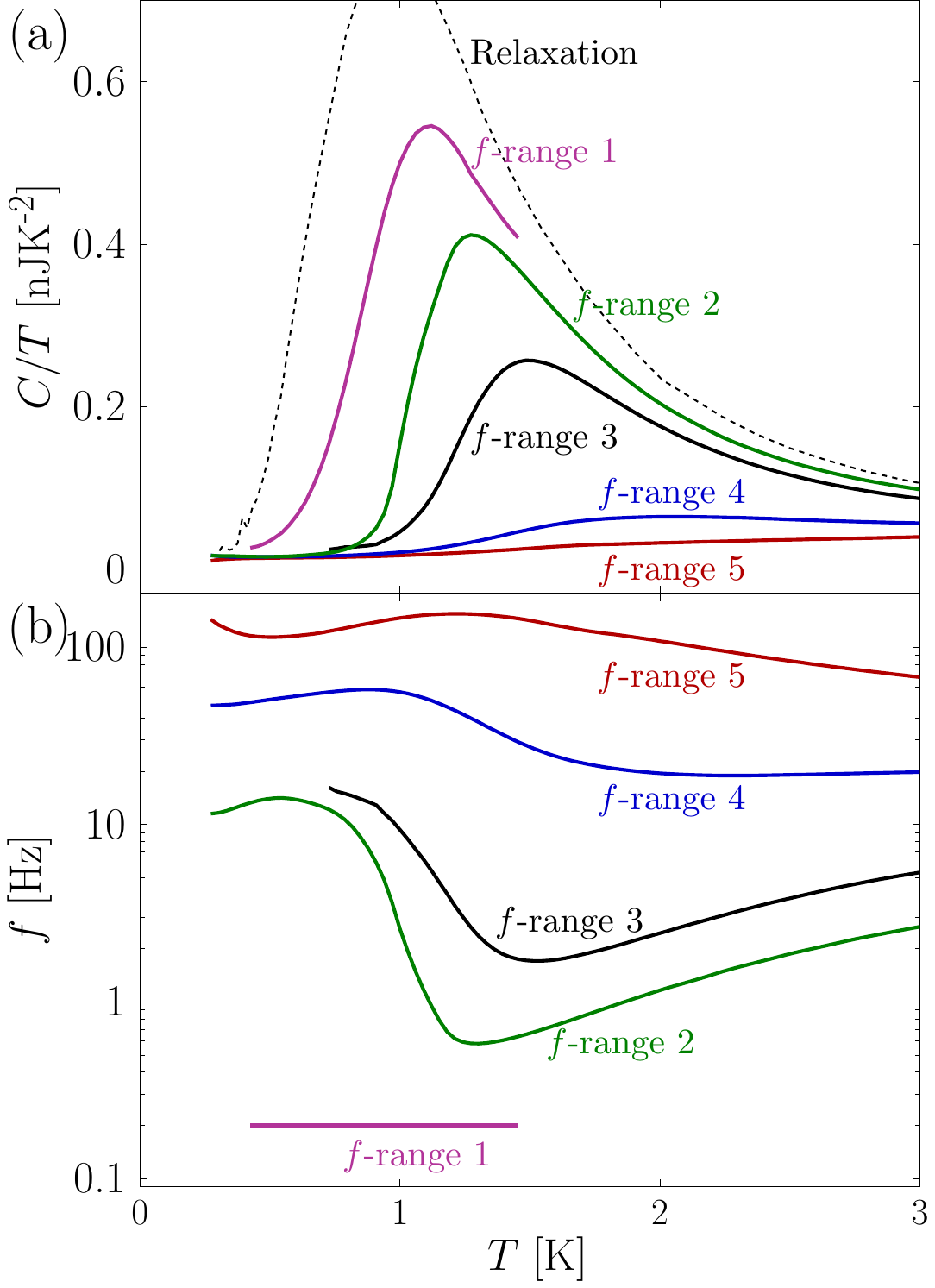}
	\caption{(a) Experimentally measured heat capacity for 30 ng powder sample of DTO mixed with Apiezon N grease. The different curves show the heat capacity at different frequencies in the 0.1\,Hz to 200\,Hz range. The dashed line shows the result from the relaxation measurement of Fig.~\ref{fig1:dtomassive} corresponding to the zero frequency limit. It has been scaled to fit the ac calorimetry measurements at $4\textup{ K}$. (b) Ac frequency as a function of temperature.}
	\label{fig2:DTO_S1}
\end{figure}

Figure~\ref{fig2:DTO_S1} establishes a clear frequency dependence in the measured dynamic heat capacity. For low modulation frequencies the dynamic heat capacity approaches the conventionally measured static value, with a characteristic spin-ice peak centered at $\sim$1\,K. When the frequency is increased, this feature is suppressed and the maximum is shifted to higher temperatures. In the limit of high frequency, the heat capacity is reminiscent of a pure phononic contribution.

\begin{figure*}[t!]
    \centering
    \includegraphics[width=0.95\linewidth]{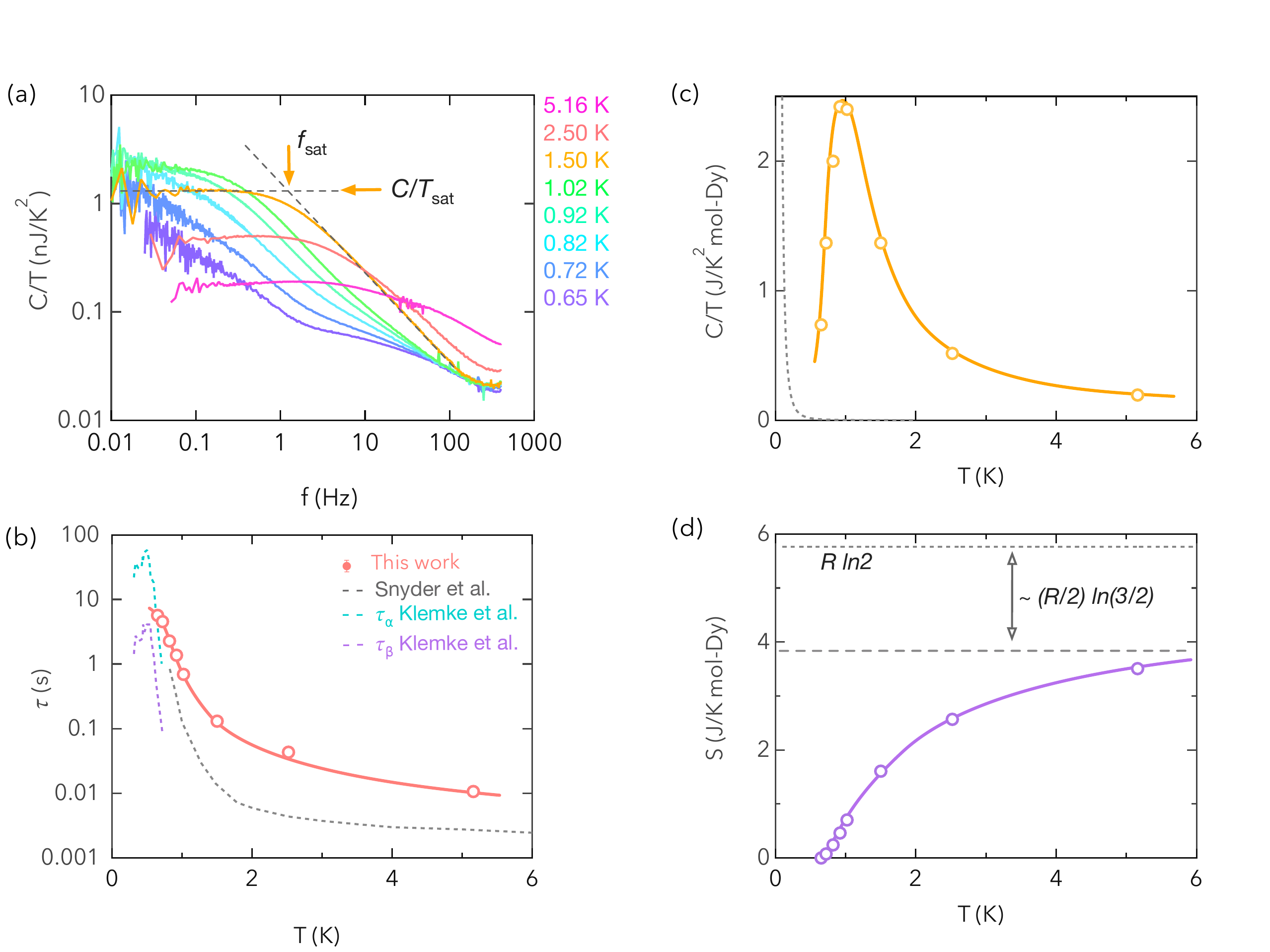}
    \caption{AC-calorimetry frequency dependence of a $\sim 250$\,ng (0.95\,nmol-Dy) DTO sample. (a) Log-log plot of frequency dependence of the dynamic heat capacity for different temperatures. The frequency (f$_{\mathrm{sat}}$) below which the measured heat capacity at 1.50\,K is independent of frequency is indicated by a vertical arrow. (b) Temperature dependence of $\tau$, taken as  1/$\omega_{\mathrm{sat}}$ where $\omega_{\mathrm{sat}} = 2\pi\text{f}_{\mathrm{sat}}$. The dashed gray curve represents the relaxation time in ac-susceptibility from Ref.~\cite{Snyder2004}, and the purple and cyan dashed curves show the relaxation time in relaxation calorimetry from Ref.~\cite{Klemke2011}. (c) Corresponding temperature dependence of the specific heat extracted from the saturation value $(C/T)_{\mathrm{sat}}$ of the frequency dependence in panel a, indicated by the horizontal arrow for 1.50\,K in panel a. The dotted curve shows the estimated nuclear specific heat (see Appendix~\ref{sec:Nuclear}). (d) The entropy from integrating the specific heat curve in panel c. The difference in the saturated value of entropy at high temperature and $R\ln{2}$ is associated with the zero-point Pauling's entropy due to the degenerate ground state configurations of the spin-ice \cite{Pauling1935}.
    The red curve in b, yellow curve in c, and purple curve in d are guides to the eye.}
    \label{fig3:TISP}
\end{figure*}

To investigate the effect of frequency on the measured heat capacity, we studied the thermal impedance \cite{Khansili2023calorimetric} and frequency dependence of the dynamic heat capacity at several different temperatures. Figure~\ref{fig3:TISP}(a) shows the measured $C/T$ for a $\sim250$\,ng DTO powder sample as a function of frequency at different temperatures. It is seen in the frequency dependence that the heat capacity saturates to a fixed value $(C/T)_{\mathrm{sat}}$ below a certain frequency. The crossover frequency is obtained by the crossing of the behaviors below and above the crossing point, indicated by the dotted lines in Fig.~\ref{fig3:TISP}(a).

Figure~\ref{fig3:TISP}(b) shows the relaxation time $\tau$ defined as 1/$\omega_{\mathrm{sat}}$, where $\omega_{\mathrm{sat}} = 2\pi \text{f}_{\mathrm{sat}}$. The relaxation time shows a divergence behavior appearing at $\sim$2\,K. This is qualitatively similar to the spin relaxation time observed in the dynamic ac-susceptibility measurements \cite{Snyder2004} and thermal relaxation \cite{Klemke2011} shown as dotted curves in Fig.~\ref{fig3:TISP}(b). The relaxation time $\tau$ reaches a value of $\sim$6\,s at 0.65\,K. The corresponding specific heat extracted from the saturation value $(C/T)_{\mathrm{sat}}$ in Fig.~\ref{fig3:TISP}(a) is shown in Fig.~\ref{fig3:TISP}(c). The specific heat obtained in such a way is similar to the one that was obtained from relaxation measurements in Fig.~\ref{fig1:dtomassive}. The nuclear specific heat in \DTO\ mainly comes from the nuclear quadrupole of Dy with nuclear spin ${I = 5/2}$ \cite{Stone2016}. The estimated nuclear quadrupole specific heat (see Appendix~\ref{sec:Nuclear}) is shown as a dotted line in Fig.~\ref{fig3:TISP}(c). For the purpose of this work, the nuclear specific heat is insignificant at the temperatures above 0.5\,K. However, for the specific heat measurements below 0.5\,K careful treatment of the nuclear specific heat is required \cite{henelius16}. 

The entropy is shown in Fig.~\ref{fig3:TISP}(d), obtained by integrating the $C/T$ curve of Fig.~\ref{fig3:TISP}(c). The entropy reaches a value close to 3.5\,J/K\,mol-Dy at 5.16\,K. The high-temperature saturation value of the entropy from Fig.~\ref{fig3:TISP}(d) is approximately $\sim$4\,J/K\,mol-Dy.
The ground state of the spin-ice is degenerate with 6 equivalent configurations that satisfy the `ice rule' where the spins have 2-in \& 2-out configuration in the tetrahedron \cite{Castelnovo_Nature}. This degenerate ground state results in the zero-point Pauling's entropy \cite{Pauling1935} of $(R/2)\ln(3/2)$ ($\approx$1.68\,J/K\,mol) per spin. Since the expected high-temperature entropy per Dy spin must be $R\ln{2}$, the missing entropy approximately reflects this ground state Pauling's entropy indicated by the vertical arrow in Fig.~\ref{fig3:TISP}(d).

The excited state of these spin configurations on the tetrahedron can be attained by flipping a single spin in the diamond bond of the tetrahedron which gives rise to a monopole-antimonopole pair \cite{Castelnovo2012, Castelnovo_Nature}. These monopole pairs are deconfined and there is no additional cost to separate them over large distances \cite{Castelnovo2012, Castelnovo_Nature}. Given this microscopic picture, the system can be effectively expressed by the dynamics of these monopole-antimonopole pairs \cite{Castelnovo2012, Castelnovo_Nature}.

\section{Theoretical analysis}
\subsection{Thermal conductivity}\label{ThermalConductivitySection}

A frequency dependence in the heat capacity measurement indicate that the system is out of equilibrium. The observed non-equilibrium contributions to the heat capacity can be due to several causes. A natural starting point is to investigate the heat transport in spin ice, which is mediated by spin and phonon interactions. The steady-state thermal conductivity of DTO has been studied previously \cite{KlemkeDTOCV,heatConductivityDTO} and it was concluded that the largest contribution to the thermal conductivity in the 1\,K range is due to phonons. By application of magnetic fields it was also found that magnetic monopoles transport heat as a smaller secondary effect.

Based on the available heat conductivity data from one of the previous studies \cite{heatConductivityDTO}, we estimate the typical response to ac calorimetry for our sample. To do this, we simulate the heat equation for a sphere with sinusoidally varying surface temperature,

\begin{equation}\label{eq:heatEquation}
\begin{split}
&	C_V(T(r,t))\frac{\partial T(r,t)}{\partial t}=\\
&r^{-2}\frac{\partial }{\partial r}\left(r^2\kappa(T(r,t))\frac{\partial T(r,t)}{\partial r}\right),
\end{split}
\end{equation}
where the thermal conductivity $\kappa$ and the specific heat $C_V\equiv C/V$ are temperature dependent with values according to experimental observations \cite{heatConductivityDTO}.  The sphere resembles the average shape of a powder grain, but the results may also be used to make order of magnitude estimates for single crystal samples. The surrounding Ag or Apiezon N grease both have high thermal conductivity in comparison to DTO and we do not need to consider their contribution for this estimate.

If the frequency is high, and the temperature gradients consequently are large,  the temperature oscillations at the center of the sample will have a smaller amplitude than those at the surface. This will result in a lower measured heat capacity since the whole sample does not participate in the heat exchange.
\begin{figure}[t!]
	\centering
	\includegraphics[width=1\linewidth]{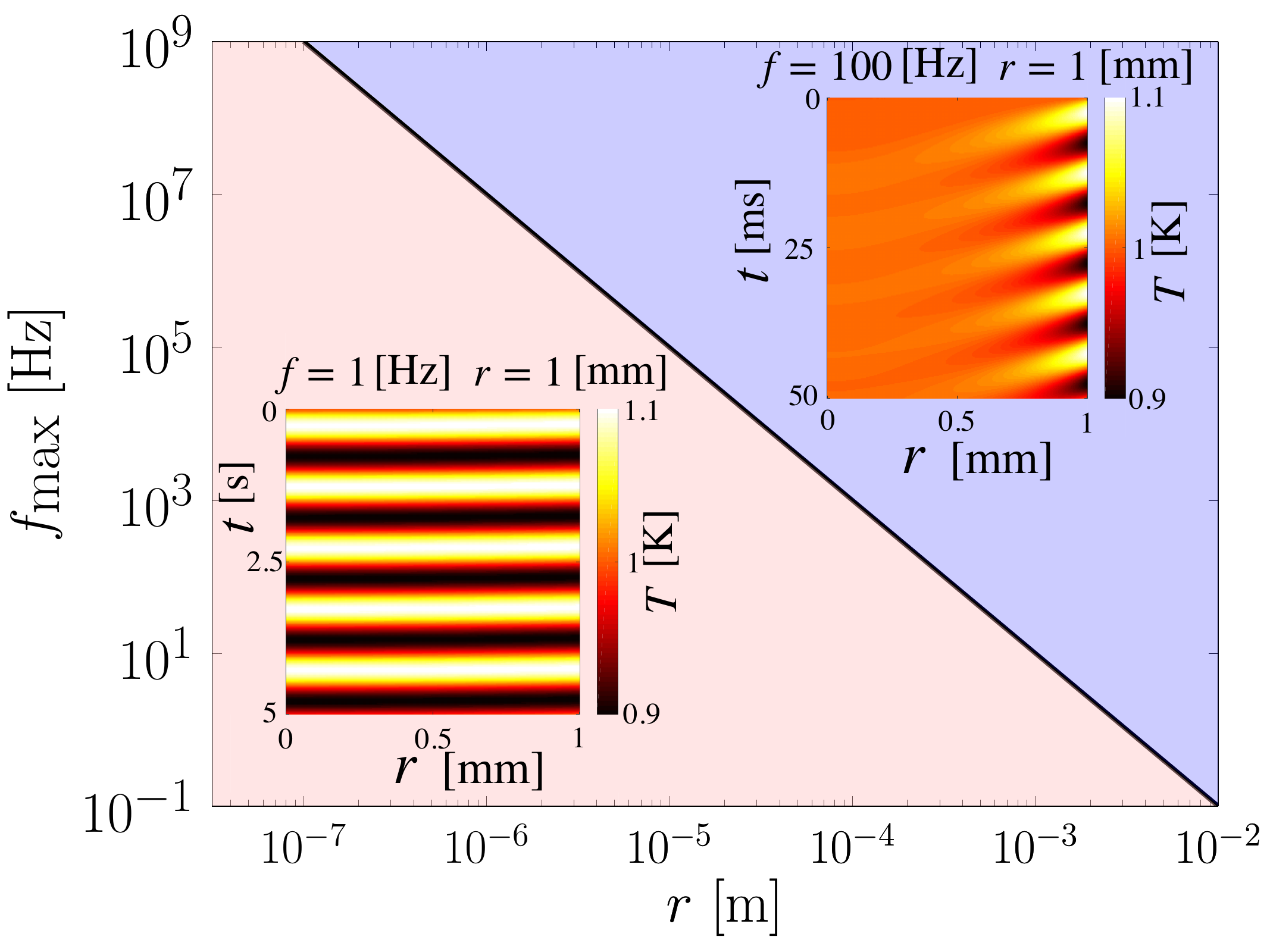}
	\caption{Theoretical upper frequency $f_\textup{max}$ that can be used to probe a spherical DTO sample of radius $r$. The upper frequency is an estimate above which the temperature oscillations in the core have an amplitude less than $10\%$ of the surface amplitude. Pink area: core and surface are at equal temperature resulting in reliable results from ac calorimetry. Blue area: surface oscillations are too fast and the core does not have time to equilibrate, resulting in a smaller measured heat capacity than the actual value. The insets show the typical solution to the heat equation, Eq.~(\ref{eq:heatEquation}), in the respective regimes for a $1\textup{ mm}$ radius sphere of DTO.}
	\label{fig4:dtothermalgradient}
\end{figure}
The heat equation allows us to determine the extent of this effect and we can determine an upper limit for the frequency that can be used to probe a sample while maintaining low thermal gradients. Figure~\ref{fig4:dtothermalgradient} shows the maximum allowed frequency $f_\textup{max}$ as a function of sample size for a spherical DTO sample of radius $r$.  The insets show the typical solutions to the heat equation in the two different regimes. Above the maximum frequency, $f_\textup{max}$, the temperature is not homogeneous in the sample, resulting in a lower value of the measured heat capacity.  For a sample with radius 1\,mm, the thermal conductivity is sufficient for frequencies up to 1\,Hz (left inset in Fig~\ref{fig4:dtothermalgradient}). However, with a surface modulation frequency of 100\,Hz the temperature oscillations affect only the surface, and the interior of the sample will not give any contribution to the measured heat capacity (right inset in Fig.~\ref{fig4:dtothermalgradient}).

If we probe below the critical ac frequency, the temperature gradients are small and we should observe the full heat capacity of the sample. This is the case for the experiments performed in this study, since the powder samples have only $\mu\textup{m}$-sized grains and the maximum frequency used is less than 500\,Hz. Hence, the observed experimental behavior is not due to poor thermal conductivity. Instead,  it originates from the slow low-temperature magnetic dynamics inherent to spin ice. With sufficiently slow dynamics a delay in the spin system temperature can develop with respect to the phonon system which should explain the observed features in the heat capacity \cite{RelaxationBook, RevModPhys.69.1}.

\subsection{Monte Carlo}
Inspired by the results in the previous section, we assume that the sample has a uniform temperature distribution and that phonon-mediated heat conduction is good. To evaluate whether the slow dynamics of the monopole excitations can give rise to the observed frequency dependence in the specific heat  we perform dynamic Monte Carlo (MC) simulations for the general dipolar spin ice Hamiltonian \cite{Yavo08,Bovo18},

\begin{equation}\label{Hamiltonian}
\begin{split}
\mathcal{H}=\;&J_1 \sum_{\langle i,j\rangle_1}\mathbf{S}_i\cdot\mathbf{S}_j+J_2 \sum_{\langle i,j\rangle_2}\mathbf{S}_i\cdot\mathbf{S}_j\\
+&J_{3a} \sum_{\langle i,j\rangle_{3a}}\mathbf{S}_i\cdot\mathbf{S}_j+ 
J_{3b} \sum_{\langle i,j\rangle_{3b}}\mathbf{S}_i\cdot\mathbf{S}_j\\
+&Da^3\sum_{i<j}\left(\frac{\mathbf{S}_i\cdot\mathbf{S}_j}{|\mathbf{r}_{ij}|^3}-3\frac{\left(\mathbf{S}_i\cdot\mathbf{r}_{ij}\right)\left(\mathbf{S}_j\cdot\mathbf{r}_{ij}\right)}{|\mathbf{r}_{ij}|^5}  \right),
\end{split}
\end{equation}
where $\textbf{S}_i$ are unit Ising spins on the pyrochlore lattice, with Ising axis along the local $\langle111\rangle$ directions. The nearest-neighbor distance is denoted by $a$ and the parameters $J_1$, $J_2$ and $J_{3a,b}$ are the first- second- and third-neighbor exchange interactions respectively. The reason for having two different types of third-neighbor interactions is that third-neighbor pairs can have two different local environments. Analogously $\langle\cdots\rangle_{i(a,b)}$ denotes summation over the $i$:th nearest-neighbor pairs. We use the {$g^+$}-DSM parameters \cite{Bovo18} $J_1=3.41\textup{ K}$, $J_2=-0.14\textup{ K}$, $J_{3a}=0.030\textup{ K}$, $J_{3b}=0.031\textup{ K}$ and $D=1.32\textup{ K}$. The simulated super-cell is cubic and built from $L^3$ standard unit cells \cite{MelkoGingras} with $L\in[1,\cdots,8]$.

The simulation is performed in the following way: the system is equilibrated at a given temperature, $T_0$, using the Metropolis-Hastings heat-bath algorithm \cite{MelkoGingras}. The process of ac calorimetry is then emulated by changing the temperature in the Boltzmann weight according to
\begin{equation}
	T=T_0+T_\textup{AC}\sin{\left(2\pi f g(T) t_\textup{MC}\right)},
\end{equation}
where $g(T)$ is a conversion factor from MC time, $t_\textup{MC}$, to real time, $f$ is the ac frequency, $T_0$ is the measured temperature and $T_{\textup{AC}}$ is the amplitude of the oscillation. A small value of $T_{\textup{AC}}$ gives the most accurate results at the cost of slow MC convergence. We find that having $T_{\textup{AC}}$ set to $1\%$ or $10\%$ of the measured temperature gives the same simulation result and set it to $10\%$ for faster convergence. Single spin-flip updates are performed at random sites in the simulation-cell with the oscillating Boltzmann weight. Every update emulates a direct spin-phonon interaction \cite{RelaxationBook}, and the Boltzmann weight reflects the interaction with a phonon-bath with oscillating temperature. The frequency dependence is emulated by letting the number of update attempts in an oscillation period be inversely proportional to the frequency.
For high frequencies, there are few updates per period. For low frequencies, there are many updates per period. This reflects that the number of spin-phonon interactions per time is constant. The temperature dependence of $g(T)$ has been introduced because there are fewer phonon modes available at low temperatures. A random collision (Monte Carlo update) therefore takes longer time to occur at low temperatures. We assume that during one second the number of spin-phonon interactions $n(T)$ follows an Arrhenius distribution,
\begin{equation}\label{eq:n_and_n0}
	n(T) = n^0 e^{-\theta/T}.
\end{equation}

The constants $n^0$ and $\theta$ are free parameters which determine the relationship between MC time and real time and need to be fitted to experimental measurements. With this assumption, the conversion factor between MC time and real time can be expressed as:
\begin{equation}\label{eq:conversionFactorgofT}
    g(T)\equiv \frac{1}{n(T)} = \frac{e^{\theta/T}}{n^0}.
\end{equation}
since there are $n(T)$ MC steps per second.

The heat capacity is calculated by fitting a linear interpolation to a $E$ vs $T$ diagram. We also add the theoretical phonon contribution according to the Debye model \cite{KlemkeDTOCV,ConceptsInThermalPhysics}:
\begin{equation}\label{eq:DebyeModel}
    C_\textup{ph}(T)= 9 n_\textup{Dy} k_\textup{B} N_\textup{A}\left(\frac{T}{\theta_D}\right)^3\cdot\int_0^{\theta_D/T}\frac{x^4e^x}{(e^x-1)^2}dx,
\end{equation}
where $n_\textup{Dy}$ = 11 is the number of atoms per molecule \DTO\ and we use the Debye temperature $\theta_D= 283\textup{ K}$ \cite{KlemkeDTOCV}. We compare the calculated heat capacity with the experiment and choose $n^0$ and $\theta$ to get the best possible fit. The size of the sample is also fitted, since it could not be determined with the needed accuracy. Hence we fit a total of three parameters to the experimental data.

\begin{figure}[t!!]
	\centering
	\includegraphics[width=0.9\linewidth]{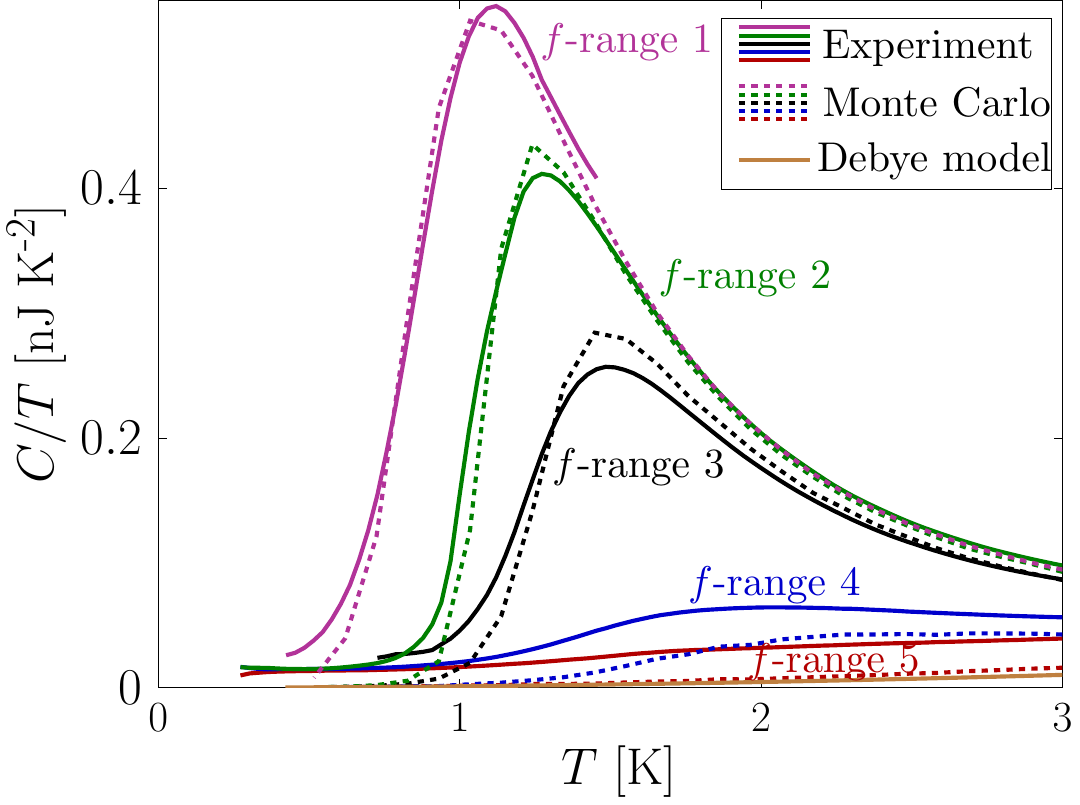}
	\caption{Theoretical MC simulation for $n  ^0=80$ and  $\theta= 2\textup{ K}$ for $L=3$ (dashed) together with experimental data (solid) for a DTO powder sample. The frequencies used for the different curves are the same as those shown in Fig.~\ref{fig2:DTO_S1}(b) ($f$-range 1-5). The contribution from the Debye model has been included in the theoretical curves, and is also plotted independently.}
	\label{fig5:acmcdtos1}
\end{figure}

\section{Discussion}
From the MC simulations in the previous section, we found that the values $n^0=80$ and $\theta=2\textup{ K}$ gave the best fit, Fig.~\ref{fig5:acmcdtos1}. From the figure, it is clear that the main features of the experiment are in excellent agreement with the simulation. This suggests that the frequency dependence in the experimental data is due to magnetic monopole relaxation effects and hence gives a new way of investigating the dynamics of monopoles in spin ice systems. 
For the higher frequencies, $f$-range 4-5, there is some discrepancy between the experiment and the MC simulation. In particular, there is a constant contribution to the experimental heat capacity which cannot be fully explained by the model. In the figure, we also show explicitly the contribution of the Debye model, Eq.~(\ref{eq:DebyeModel}), which has been added to all MC curves. For $f$-range 5 the theoretical heat capacity is almost only due to this contribution. That is: the simulation is too fast to capture any magnetic contribution.

Compared to other techniques, the value $\theta=2\textup{ K}$ is in the same order of magnitude as the previously reported value of $4.5\textup{ K}$ \cite{Revell13}, found by susceptibility measurements. Furthermore, inserting $n^0$ and $\theta$ in $g(T)$ we find that one MC step at 1\,K corresponds to about $100\textup{ ms}$. By extrapolation to $T=4\textup{ K}$ we obtain $20\textup{ ms}$ per MC step in contrast to $2.5\textup{ ms}$  mentioned in previous studies \cite{JaubertMCtime}. The discrepancy is worth noting and can have several causes. The previous study uses a different technique for studying the relaxation phenomena, which can have resulted in a timescale characteristic for that experiment. Another possible cause of the discrepancy is an oversimplification in treating MC time like real time. Nevertheless, the excellent fit with experimental data indicates that the approximation can still give a good description of the underlying physical properties.

\section{Conclusion}

Many recent experimental studies have aimed to recover the \textit{lost} Pauling entropy and to find signs of the associated predicted peak \cite{Melko01} in the specific heat originating  in a low-temperature phase transition at about 0.2\,K. We have investigated the low-temperature specific heat in \DTO. We find that ac calorimetry in the frequency range of 0.01-500 Hz is an experimental probe suitable for studying thermal relaxation effects in classical spin ice compounds such as DTO. Within this frequency range the spin ice specific heat peak, centered around 1\,K, changes from almost fully developed below 0.1 Hz to a pure phonon contribution above 200 Hz. 

The relaxation time extracted from the frequency dependence of the specific heat shows similar qualitative behavior as the spin relaxation time in ac-susceptibility \cite{Snyder2004} and previous thermal relaxation measurements \cite{Klemke2011}. The spin relaxation time $\tau$ shows a divergence behavior below $\sim$2\,K with $\tau$ reaching about $\sim$6\,s at 0.65\,K. Using Monte-Carlo simulations we show that this dynamical response of specific heat comes directly from the slow dynamics of the monopoles. This shows that ac calorimetry is an effective experimental probe for investigating slow-dynamical spin systems.
However, in this study, we have not been able to push the low-temperature limit of a well-equilibrated calorimetric measurements down to 250 mK, which most likely would be needed to resolve the outstanding question about an ultimate low-temperature ordered spin ice phase in DTO. Due to the effects of a number of likely physical phenomena such as material disorder, extremely slow intrinsic dynamics, quantum effects and a refrustration of the long range order this remains an experimental as well as theoretical grand challenge.

\begin{acknowledgments}
The simulations were performed on resources provided by the Swedish National Infrastructure for Computing (SNIC) at the Center for High Performance Computing (PDC) at the Royal Institute of Technology (KTH). The project was supported by Nordforsk through the program NNSP (Project No. 82248) and by the Danish Agency for Research and Innovation through DanScatt. A.K. and A.R. acknowledge support by the Swedish Research Council, Grant No. 2021-04360.
\end{acknowledgments}

\appendix

\section{Thermal Relaxation}
\label{sec:ThermalRelax}
In order to confirm that the materials are correctly synthesized and that the experimental setup is working with the desired precision for spin-ice compounds, we first perform thermal relaxation measurements. Powder of DTO is cold pressed with silver powder and cut into a block of size approximately $800\times 800\times 500~\mu \textup{m}$. The relaxation is performed by decreasing the applied power so that the temperature drops with about 50 steps per decade from 4\,K to 0.2\,K.
The power $P_{\textup{out}}(T)$ escaping through the thermal link is estimated from the equilibrium temperature at each step. We interpolate over these values, and find the heat capacity $C(T)$ by minimizing the residual between the solution of
\begin{equation}\label{eq:heatCapacityODE}
	\frac{dT}{dt} = \frac{P_\textup{in}(t)-P_\textup{out}(T(t))}{C(T)},
\end{equation}
and the experimental data. The heat capacity is assumed constant for each relaxation interval and $P_\textup{in}(t)$ is the power applied via the heater.

\section{AC Calorimetry}
\label{sec:ModulationCal}
The power $P(t)$ is modulated sinusoidallly with frequency $f=\omega/2\pi$, resulting in a temperature oscillation  $T_\textup{ac}(t)$ of the sample \cite{TAGLIATI201166}
\begin{equation}\label{eq:differentialEquationToDescribeModulationCalorimetry}
	\begin{split}
		&P(t)=P_0(1+\sin{\omega t}),\\
		&T_\textup{ac}(t) = T_\textup{ac,0}\sin(\omega t -\phi).\\
	\end{split}
\end{equation}
Here, $P_0$ is a constant and by varying the input power and the modulation frequency, the amplitude of the temperature oscillations $T_\textup{ac,0}$ and its phase shift $\phi$ is set depending on the properties of the sample. By measuring these quantities, the heat capacity $C$ can be computed according to \cite{GMELIN19971}
\begin{equation}\label{eq:solutionOfDifferentialEquatioForModulationCalorimetry}
	\begin{split}
		&T_\textup{ac,0} = \frac{P_0}{\sqrt{(\omega C)^2 + \kappa^2}},\\
		&\tan \phi = \frac{\omega C}{\kappa},
	\end{split}
\end{equation}
where $\kappa$ is calorimeter-bath thermal conductivity.

For this measurement technique, it is advantageous \cite{TAGLIATI201166} to adjust the frequency such that the phase shift between power and temperature, $\phi$, is kept constant. The frequency used therefore varies with temperature as is presented in Fig.~\ref{fig2:DTO_S1}(b), where each curve, $f$-range 1-5, corresponds to a preset $\phi$. The temperature oscillation due to the AC heater power is always a constant fraction of $T$ as well. 

\section{Nuclear Specific Heat}
\label{sec:Nuclear}

The nuclear quadrupole Hamiltonian \cite{Abragam1961, Slichter2013} is 
\begin{align}\label{eq:nuclearquadrupole}
    H_Q = \frac{e^2qQ}{4I(2I-1)} (3I_z^2 - I^2)
\end{align}
where $Q$ is quadrupole moment, $e$ is the electric charge on the nucleus, $q$ is the electric field gradient and $I$ is the nuclear spin.

For Dy (nuclear spin $I = 5/2$), in the limit of large temperature, $T \gg e^2qQ/4I(2I-1)$, the nuclear quadrupole specific heat is expressed as

\begin{align}\label{eq:nuclearspecificheat}
    C_Q = \frac{7R}{200} \Big(\frac{e^2qQ}{k_B T}\Big)^2.
\end{align}

Since the electric field gradient is not known for Dy in \DTO\ (which could be significantly different from Pure Dy), we use the values for pure Dy from Ref.~\cite{Anderson1969} to estimate the nuclear specific heat. 

The isotopes of Dy, which have non-zero nuclear spin $I$, are $^{161}$Dy and $^{163}$Dy, both of which have $I = 5/2$. They have natural abundances of 18.9\% and 24.9\%, respectively. For pure Dy, the value of parameter $P = \frac{3}{k_B} ( \frac{e^2qQ}{4I(2I-1)}) = \frac{3}{40} \frac{e^2qQ}{k_B} $ is 0.0093\,K for $^{161}$Dy and 0.0098\,K for $^{163}$Dy. we obtain the factor $e^2qQ/k_B$ as 0.124\,K and 0.130\,K for $^{161}$Dy and $^{163}$Dy, respectively.

Plugging in these values to Eq.~(\ref{eq:nuclearspecificheat}) we get the nuclear specific heat ${^{161}C_Q = 0.0044/T^2}$ J/K mol-Dy and  ${^{163}C_Q = 0.0049/T^2}$ J/K mol-Dy. Thus, the total nuclear quadrupole-specific heat is 

\begin{align}
    ^{Dy}C_Q  = (0.189~ ^{161}C_Q + 0.249~ ^{163}C_Q) = 0.00205/T^2
\end{align}

Even though the nuclear contribution is small for the values provided from the pure Dy, the actual values might differ significantly for Dy in \DTO. Since the electric field gradient `$q$' at the atomic site is strongly dependent on the chemical environment, an NMR/NQR study is required to extract the electric field gradient in \DTO\, to estimate the exact nuclear contribution in specific heat.

\bibliographystyle{apsrev4-1}
\bibliography{references}

\end{document}